\documentclass[10pt,twocolumn,preprintnumbers,amsmath,amssymb,nofootinbib
,superscriptaddress]{revtex4-1}
\usepackage{graphicx,longtable,mathrsfs,color,array}
\usepackage[hidelinks]{hyperref}
\usepackage[usenames,dvipsnames]{xcolor} 
\usepackage{amssymb,amsmath,mathtools,mathrsfs,slashed} 
\usepackage{epsfig,subfigure,placeins,float} 
\usepackage{booktabs,longtable,ctable,multirow} 
\usepackage{exscale,relsize} 
\usepackage[normalem]{ulem} 
\usepackage{enumerate}
\usepackage{comment}
\usepackage{color}
\allowdisplaybreaks[1]

 \definecolor{DarkBlue}{cmyk}{1,1,0,0.1}

\begin{document}
\title{Apparent Superluminality of Lensed Gravitational Waves}
\author{Jose Mar\'ia Ezquiaga}
\email{NASA Einstein fellow; ezquiaga@uchicago.edu}
\affiliation{Kavli Institute for Cosmological Physics, Enrico Fermi Institute, The University of Chicago, Chicago, IL 60637, USA}
\author{Wayne Hu}
\email{whu@background.uchicago.edu}
\affiliation{Kavli Institute for Cosmological Physics, Enrico Fermi Institute, The University of Chicago, Chicago, IL 60637, USA}
\affiliation{Department  of  Astronomy \& Astrophysics, The University of Chicago, Chicago, IL 60637, USA}
\author{Macarena Lagos}
\email{mlagos@kicp.uchicago.edu}
\affiliation{Kavli Institute for Cosmological Physics, Enrico Fermi Institute, The University of Chicago, Chicago, IL 60637, USA}
\date{Received \today; published -- 00, 0000}

\begin{abstract}
We consider gravitational wave (GW) sources with an associated electromagnetic (EM) counterpart, and analyze the time delay between both signals in the presence of lensing. If GWs have wavelengths comparable to the Schwarzschild radius of astrophysical lenses, they must be treated with wave optics, whereas EM waves are typically well within the approximation of geometric optics. With concrete examples, we confirm that the GW signal never arrives before its EM counterpart, if both are emitted at the same time. However, during the inspiral of a binary, peaks of the GW waveform can arrive before their EM counterpart. 
We stress this is only an apparent superluminality since the GW waveform is both distorted and further delayed with respect to light. 
In any case, measuring the multi-messenger time delay and correctly interpreting it has important implications for unveiling the distribution of lenses, testing the nature of gravity, and probing the cosmological expansion history.
\end{abstract}

\date{\today}
\maketitle


\section{Introduction}\label{sec:introduction}
Multi-messenger gravitational wave (GW) astronomy has arrived to revolutionize our understanding of extreme, astrophysical phenomena as well as shedding new light on different aspects of  cosmology  and  fundamental  physics. 
For instance, the time delay between GW170817, the first binary neutron star detected by the LIGO/Virgo collaboration \cite{TheLIGOScientific:2017qsa}, and its associated electromagnetic (EM) counterpart \cite{2041-8205-848-2-L14, 2041-8205-848-2-L15} served to set an impressive constraint on the propagation speed of GW with respect to the speed of light, $|c_\mathrm{gw}/c-1|\lesssim 10^{-15}$; the graviton mass, $m_g<9.51\cdot 10^{-22}{\rm eV}/c^2$ \cite{Abbott:2018lct}; and, as a consequence, many dark energy models \cite{2017PhRvL.119y1301B, Ezquiaga:2017ekz, Creminelli:2017sry, 2017arXiv171005893S}, and dark matter alternatives \cite{Boran:2017rdn}. 
Still, GW170817 was a relatively nearby transient, $z\approx 0.01$. 
Future GW detections may have much higher redshifts, raising the probability of being lensed by inhomogeneities in the universe along the line of sight. Since gravitational lensing introduces additional time delays, it is crucial to understand those first before interpreting anything about cosmology or the theory of gravity. 

Lensing of waves can be treated using the geometric optics approximation when the wavelength of the signal is small compared to the Schwarzschild radius of the lens: $\lambda \ll R_s$ \cite{1998PhRvL..80.1138N, Takahashi:2003ix, Macquart:2004sh,Takahashi:2016jom}, otherwise interference becomes relevant and one must use wave optics. 
We thus see that for any
GW experiment operating in a given GW frequency band, e.g. LIGO \cite{2015CQGra..32g4001L}, LISA
\cite{2017arXiv170200786A}, IPTA \cite{2010CQGra..27h4013H,2013CQGra..30v4010M},
wave effects can be important if the emission is lensed by a sufficiently small mass.

The multi-messenger time delay taking into account interference effects for GWs has been studied before. Different definitions have been made, using either the phase~\cite{Takahashi:2016jom}, group \cite{Morita:2019sau} or front velocity \cite{Suyama:2020lbf} of the wave, arriving to contradicting results. In particular, \cite{Takahashi:2016jom, Morita:2019sau} argue that GWs can arrive before its simultaneously emitted EM counterparts, whereas \cite{Suyama:2020lbf} shows that if GWs arrive before light then there would be a violation of causality (since light in the geometric optics regime takes the shortest path from the source to the observer), and therefore GWs can never arrive before light in the context of General Relativity. 

In this paper, we revisit the multi-messenger time delay in the presence of lensing, and 
resolve the apparent contradiction between previous studies. In particular, we clarify the meaning of the different definitions of time delay. By considering both simple and realistic examples, we show that certain features in the GW waveform are determined by the phase and group peaks and may indeed appear to arrive in advance to the corresponding feature in the EM signal. However, simultaneously turned on GW and EM signals will arrive at the same time, since the front velocity of GWs coincides with that of light. 
We thus conclude that there is no superluminal propagation of GWs. Instead, interference and amplification along multiple paths distorts the GW signal and this can cause the phase peaks of the signal to appear in advance.  This apparent superluminality of the phase time delay does  not  represent  a  signal  that  actually  propagates  superluminally, although there can still be observational consequences if there is an EM counterpart that also carries this phase information. Analogous arguments hold for the group time delay.

This paper is organized as follows. In Section \ref{sec:timedelay} we define the various time delays between GW and EM waves. We employ the geometric approximation for EM waves, and wave optics for GW. We consider a point lens as an example, and discuss the different definitions of time delay, and their behaviour.
In Section \ref{sec:gaussian} we consider a toy situation of a low-frequency sinusoid signal modulated by a Gaussian. In this case we compare the arrival time of GW to light, and discuss the apparent early arrival of the GW signal. In Section \ref{sec:waveform} we study the lensing of a realistic GW emission from a binary black hole, from inspiral to ringdown. Finally, in Section \ref{sec:discussion} we summarize our findings and discuss their consequences for multi-messenger scenarios. 

\section{Gravitational Lensing Time Delay}\label{sec:timedelay}

\subsection{Time Delays}

\begin{figure}[t!]
\centering
\includegraphics[width = 0.95\columnwidth]{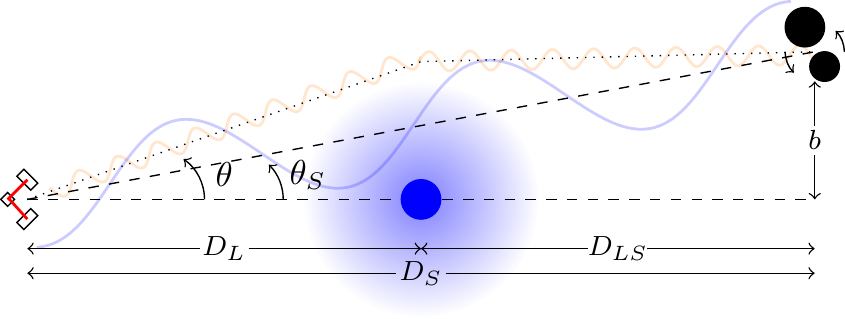}
 \caption{Diagram of lensing scenario. The observer (detector configuration, left) receives emission from  a GW source (compact binary, right), through a lens (blue circle, middle).  
  Short-wavelength (yellow) emission propagates on a path with minimum time delay whereas long-wavelength emission (blue)
 effectively propagates through multiple paths
 leading to interference and a non-vanishing time delay between the two for signals emitted at the same time. }
 \label{fig:diagram}
\end{figure}

Let us consider the lensing scenario illustrated in Fig.~\ref{fig:diagram}. Here, $D_L$, $D_S$ and $D_{LS}$ correspond to the comoving angular diameter distances from the observer to the lens, from the observer to the source, and from the lens to source, respectively. Also, $\vec{\theta}_S$ is the angular position of the source with respect to the lens, in the absence of the lens. Due to lensing, a source located
at an unlensed angular position $\vec{\theta}_S$ 
will have its lensed positions
$\vec{\theta}$ shifted due to all of the possible deflected paths around the lens.
For a given $\vec{\theta}$, the time delay between the lensed and unlensed path in the thin lens approximation is determined by:
\begin{equation}\label{Eqtd}
    t_d(\vec{\theta},\vec{\theta}_S)\approx \frac{1}{c}\frac{D_LD_S}{2D_{LS}} |\vec{\theta}-\vec{\theta}_S|^2 + t_{ \Phi}(\vec{\theta}).
\end{equation}
The first piece is the geometric time delay and $t_{\Phi}$ is the Shapiro time delay from the time
dilation due to traversing a lensed path $s$ through the gravitational potential $\Phi$:
\begin{equation}
   t_\Phi \approx -\frac{2 }{c^3} \int \Phi ds.
\end{equation}
In geometric optics, we define the images and their associated time delays by finding the stationary trajectories around which $\delta t_d=0$ in accordance with Fermat's principle. In particular, one of the images corresponds to the minimum $t_d$ and so we would expect that no GW signal can arrive earlier than high-frequency EM signals, even in the wave optics limit.

In wave optics, the wavelength is larger than the path difference between the multiple possible images, and therefore interference becomes relevant. In that case, there will be only one image formed from the superposition of all paths. One can define the amplification factor $F$ that multiplies the wave in Fourier space  and takes into account all possible paths arriving to the observer. 
 In particular, in the thin lens approximation, it is given by \cite{Schneider:1992}:
\begin{equation}
    F(\omega, \vec{\theta}_s)=\frac{D_L D_S}{D_{LS}}\frac{1}{c}\frac{\omega}{2\pi i}\int d^2\theta e^{i\omega t_d (\vec{\theta},\vec{\theta}_S)},
\end{equation}
where $\omega$ is the angular frequency of the lensed wave. 
The temporal profile of the lensed wave at the observer position will include all possible frequency components and will thus be described by:
\begin{equation}
    S_L(t, \vec{\theta}_S)=\int \frac{d\omega}{2\pi} e^{-i\omega t}F(\omega,\vec{\theta}_S)S_\omega,
\end{equation}
where we have ignored the polarization of the wave, and thus this expression can be applied to both gravitational and electromagnetic waves. Here, $S_\omega=\int dt e^{i \omega t} S(t)$
is the Fourier transform of the unlensed wave source $S(t)$, with time referenced to the detection epoch.

In wave optics, the time delay has been defined in different ways in the literature, using either the phase \cite{Takahashi:2016jom}, group \cite{Morita:2019sau} or front velocity \cite{Suyama:2020lbf} of the wave. Depending on the particular feature we are interested in for the GW signal, the time delay will be given by one or another of these delays.

The front velocity is associated with the causal propagation of a signal that
 is suddenly turned on.  Since the front is effectively discontinuous it is defined by high frequency modes and thus, by construction, the GW front will arrive at the same time as light (i.e.~both arrival times will be given by the geometric optics time delay $t_d$). Therefore, we expect no time delay between light and  GWs emitted at the same time. For example, we would not know from GWs that a merger of compact binaries occurred before light. This intuitive description agrees with the results in \cite{Suyama:2020lbf}.

We can also define a time delay $t_p$ associated to the net phase shift $\Delta \phi \equiv \omega t_p$ of a monochromatic wave as:
\begin{equation}
    t_{p}(\omega, \vec{\theta}_S)=- \frac{i}{\omega}\ln \left(\frac{F(\omega, \vec{\theta}_S)}{|F(\omega, \vec{\theta}_S)|}\right).
    \label{eq:tphase}
\end{equation}
Notice that $t_p$ is not the average of $t_d$ over all possible paths, but rather tracks the phase of the average amplification.
This time delay  therefore characterizes  the arrival time of peaks and troughs of a monochromatic, or temporally infinite plane wave, in the
wave optics regime.   As we shall see, it does not directly 
represent a time delay of a signal with a definite emission
and arrival time.
We emphasize that the amplification factor is defined in such a way that for $\omega \rightarrow 0$, where the long wavelength prevents $F$ from being dominated by paths close to the lens, $F \rightarrow 1$ and
there is no lensing amplification at all.
The phase delay $t_p$ as a fraction of
the period also goes to zero $t_p \omega \rightarrow 0$, even though
its absolute value can grow due to the
definition in Eq.~(\ref{eq:tphase}).

Since the phase shift is generically frequency dependent, it will lead to distortions of wavepackets containing multiple frequencies. In this case, the features of the overall wavepacket, which in principle can be temporally localized and carry a signal, are expected to have arrival times given by the delay of the stationary phase point, in analogy to the group velocity:
\begin{equation}
    t_{g}(\omega, \vec{\theta}_S) = t_{p}(\omega, \vec{\theta}_S)+\omega \frac{\partial t_{p}(\omega, \vec{\theta}_S)}{\partial \omega},
\end{equation}
which corresponds to setting $\partial (\phi+\Delta\phi)/\partial\omega=0$.  We shall see that this also does not necessarily correspond to the delay in the arrival time of a signal given that lensing also distorts
the group waveform.

In what follows, it will be useful to work in a set of dimensionless units for a lens of mass $M$ at redshift $z_L$:  angles in units of the Einstein ring radius
\begin{equation}
\theta_E = \sqrt{ \frac{4 G M}{c^2 } \frac{(1+z_L)D_{LS}}{D_L D_S}},
\end{equation}
and  time delays in units of the dilated Schwarzschild diameter crossing time 
\begin{equation}
t_M=4GM(1+z_L)/c^3.
\end{equation}
As a rule of thumb, it is useful to remember that for $1M_\odot$ at $z_L=0$, $t_M\approx 20\mu$s. With these units in hand we can define the following dimensionless parameters:
\begin{equation}\vec{x}=\frac{\vec{\theta}}{\theta_E}, \quad \vec{y}=\frac{\vec{\theta}_S}{\theta_E}, \quad 
w=t_{M}\omega;
\end{equation} and the dimensionless time delay:
\begin{equation}
    T_{*}=\frac{t_{*}}{t_{M}},
\end{equation} (for any arrival time, e.g.\ $* \in \{ d,p,g \}$).
Conveniently, the dimensionless frequency can also be written in terms of the Schwarzschild radius of the lens
$R_s$ and the redshifted wavelength of the signal $|w|=4\pi R_s(1+z_L)/\lambda$, so that the condition to be in the wave-optics limits can be simply stated as
\begin{equation}
    |w|<2\pi\,.
\end{equation}

Finally, in our convention, time delays are
defined such that $T_*=0$ in the absence
of lensing. If $T_*>0$, then lensed signals arrive after the unlensed signal.
Note that this  absolute sense of
time delay is often not preserved in the literature 
\cite{1974PhRvD...9.2207P,Takahashi:2016jom,Suyama:2020lbf} since only relative time delays are observable.  
In order to avoid confusion, we will express our results in terms of the multi-messenger time delay
\begin{equation}
\Delta T_*= T_{*}-T_{+}\,,
\label{eq:DeltaT}
\end{equation}
where $T_*$ will describe the GW time delay, which can be given by $T_p$ or $T_g$, whereas $T_+$ will describe the EM time delay of the first image (the one with a minimum arrival time) in the geometric optics limit.
Note that any common offsets in $T_*$ will cancel out and become irrelevant for the physical interpretation of the results. 

\subsection{Point Lens}\label{sec:examples}
For illustrative purposes, we consider a point lens of mass $M$. 
In the Born approximation, we take the lensed path
$s$ as an undeflected path in the line of sight direction $r$ with impact parameter $b$ from the lens in comoving coordinates. The lens plane corresponds to $r=0$.
In this case, we can evaluate the Shapiro time delay as
\begin{align}
    t_\Phi &\approx \frac{ t_M}{2} \int_{-D_L}^{D_{LS}} \!\! \frac{dr}{\sqrt{ r^2 + b^2}} 
    = -\frac{t_M}{2}  \ln \frac{ \sqrt{D_{LS}^2 + b^2} -D_{LS}}{\sqrt{D_L^2 + b^2}+D_L}.
    \label{eq:Shapiropoint}
\end{align}
For $\theta\ll 1$, we can approximate $b\approx \theta D_L$ and expand
\begin{equation}
    t_\Phi \approx -\frac{t_M}{2} \left[  \ln \theta^2 -\ln \frac{4D_{LS}}{D_L} \right] \approx -t_M \ln \theta.
    \label{eq:Shapiroapprox}
    \end{equation}
In the second approximation we have assumed that  $D_L$ and $D_{LS}$ remain finite and comparable, and so
the time delay is dominated by the $\ln \theta$ factor for
$\theta\ll 1$.   The
approximation~(\ref{eq:Shapiroapprox}) should not be extrapolated to $\theta \gtrsim 1$, where it would break down and predict negative time delays while
$t_\Phi$ is always positive in Eq.~(\ref{eq:Shapiropoint}).

Including the geometric term, the dimensionless time delay becomes
\begin{equation}
T_d = \frac{1}{2} \left[ (x \cos\alpha-y)^2 + x^2\sin^2\alpha) \right]- \ln x - \ln \theta_E,
\label{eq:Tdpointmass}
\end{equation}
where recall $\vec{x} =\vec{\theta}/\theta_E$, $\vec{y}=\vec{\theta}_S/\theta_E$ and we have oriented the
source at $\alpha=0$ in the polar coordinates $ (x,\alpha)$ of the 
transverse plane. 

The amplification factor then becomes
\begin{align} \label{eq:ampl_point}
F(w,\vec{y}) &= \frac{w}{2\pi i} \int_0^\infty x dx \, \int_0^{2\pi} d\alpha e^{i w T_d}  \nonumber\\
&= \frac{w}{ i}e^{-i w\ln \theta_E}  \int_0^\infty x dx  e^{i w(x^2+y^2)/2}  x^{-iw} J_0(x w y)\nonumber\\
&=
\nu^{1-\nu}e^{2\nu \ln \theta_E}\Gamma\left(\nu\right)L_{\nu}\left(-\nu\,y^2\right),
\end{align}
where $\nu=-iw/2$ and $L_n(z)$ is the Laguerre polynomial. 

Note that this expression is equivalent to the standard one quoted in the literature in terms of hypergeometric functions \cite{1974PhRvD...9.2207P}, but we have explicitly included the global phase factor $\ln\theta_E$ which is normally omitted when working in dimensionless units since it is common to all paths
(see discussion below Eq.~(\ref{eq:DeltaT})).

In the geometric optics regime, images will form at the stationary points of the time delay $T_d$:
\begin{equation}
x_{\pm}= \frac{1}{2}\left| y \pm \sqrt{y^2+4}\right|,
\end{equation}
and are aligned along the source-lens axis with $\alpha=0,\pi$ for $+$,$-$ respectively.
The time delay of light will then be given given by
evaluating Eq.~(\ref{eq:Tdpointmass}) at the image positions:
\begin{equation}
     T_\pm(y)= \frac{1}{4}\left[y^2+2\mp y\sqrt{y^2+4}\right]-\ln\,x_{\pm} -\ln\,\theta_E,
\end{equation}
where we have included a common delay for the two images $(-\ln\,\theta_E)$ typically neglected in the literature.
This factor is only important for the interpretation of the time delay with respect to a hypothetical unlensed signal.
The same offset will arise for the GW time delay $T_{*}$ from the global phase factor and thus will not affect the multi-messenger time delay $\Delta T_*$ of Eq.~(\ref{eq:DeltaT}).
Note, however, that the total $F$ function encodes the information on all the possible images, and therefore $T_p$ and $T_g$ describe delays for the total waveform, whereas the EM time delay $T_{+}$ describes the time delay of one EM image only. As a result, $\Delta T_p$ will not vanish for a point lens in the high frequency limit. 
Nevertheless, in that limit, the first arrival of GWs will coincide with that of a simultaneously emitted EM signal. 
Moreover, even in the geometric optics limit, where the amplification factor (\ref{eq:ampl_point}) can be decomposed into contributions around the $+$ and $-$  paths in the stationary phase approximation
\cite{10.1143/PTPS.133.137} (valid when $|w(T_{+}-T_{-})|\gg 1 $), the phase shifts of the two images are $w T_{+}$ and
$\left[w T_{-}-\text{sign}(w)\cdot\pi/2\right]$,
as we have explicitly verified by
summing the two phase shifted signals, weighted by $\sqrt{|\mu_{\pm}|}$, where 
$\mu_{\pm}$ are the magnifications
from geometric optics.  The appearance of $\text{sign}(w)$ ensures that the effect is a pure phase shift for a real monochromatic wave.
Therefore even in the
geometric optics limit, the phase delay cannot be directly associated with
time delay
 because it also carries information from the interference of various lensed paths around the stationary point.  Notice that the additional phase shift does not depend on the value of the frequency in the geometric optics limit and so does not change the group arrival time from $T_{\pm}$.

\begin{figure}[t!]
\centering
\includegraphics[width = 1\columnwidth]{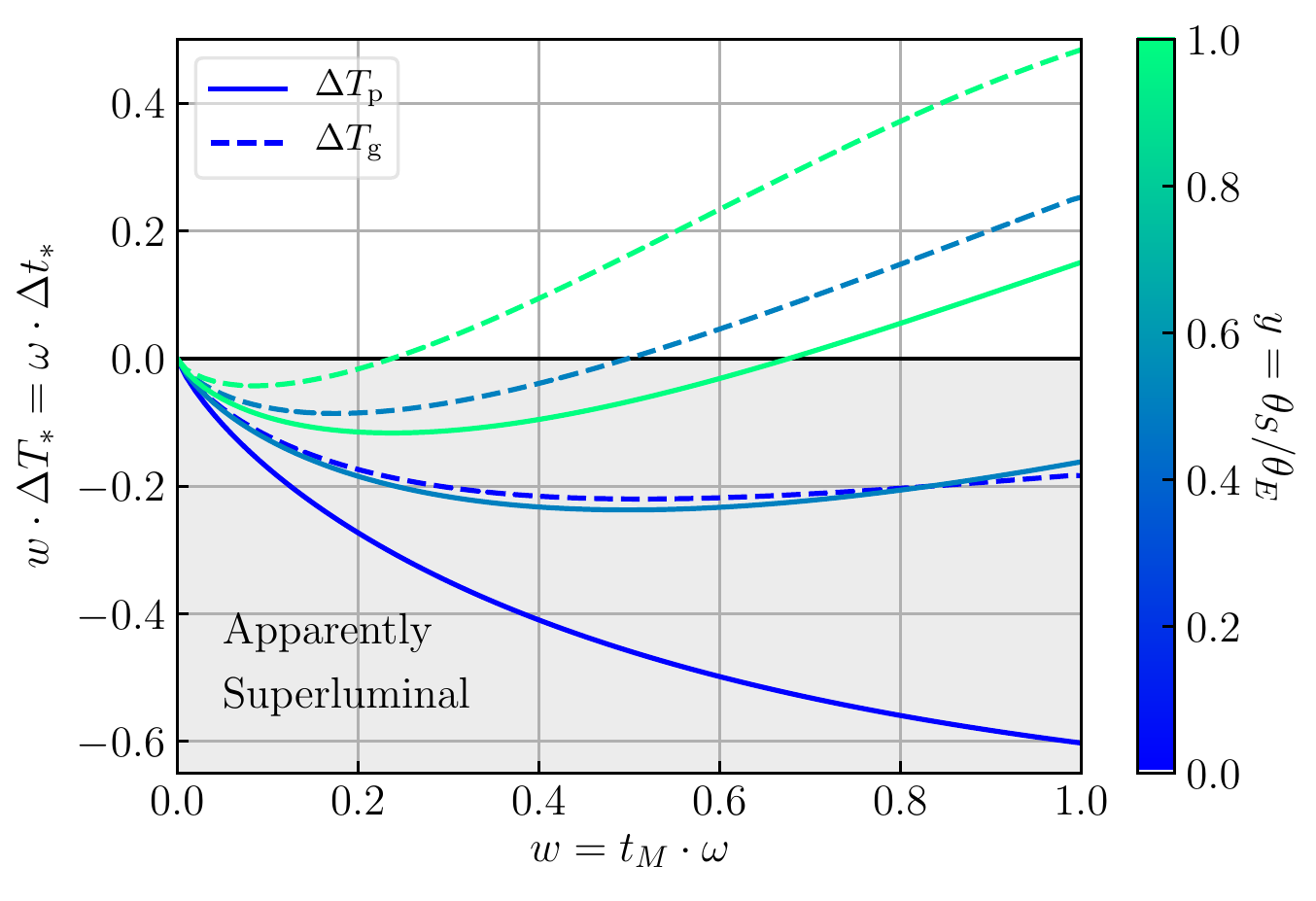}
 \caption{Difference in the phase  delay $\Delta T_{p}$ (solid lines) and group delay $\Delta T_g$ (dashed lines) with respect to
  the first arrival of light in geometric optics, in units of the Schwarzschild diameter crossing time $t_M$ and as a function of the frequency $\omega$, for three different source angles with respect to the Einstein ring radius $y=
 0,0.5,1$ of a point mass lens. $\Delta T_*<0$ corresponds to cases with an apparent arrival of the GW signal before light (shaded region).  We display arrival time delays as $w \cdot \Delta T_*$ to 
 reflect changes relative to the period, e.g.~the phase shift $\omega\cdot 
\Delta t_p$, which vanishes as $\omega\rightarrow 0$. }
 \label{fig:phase}
\end{figure}

\begin{figure*}[t!]
\centering
\includegraphics[width = 0.98\textwidth]{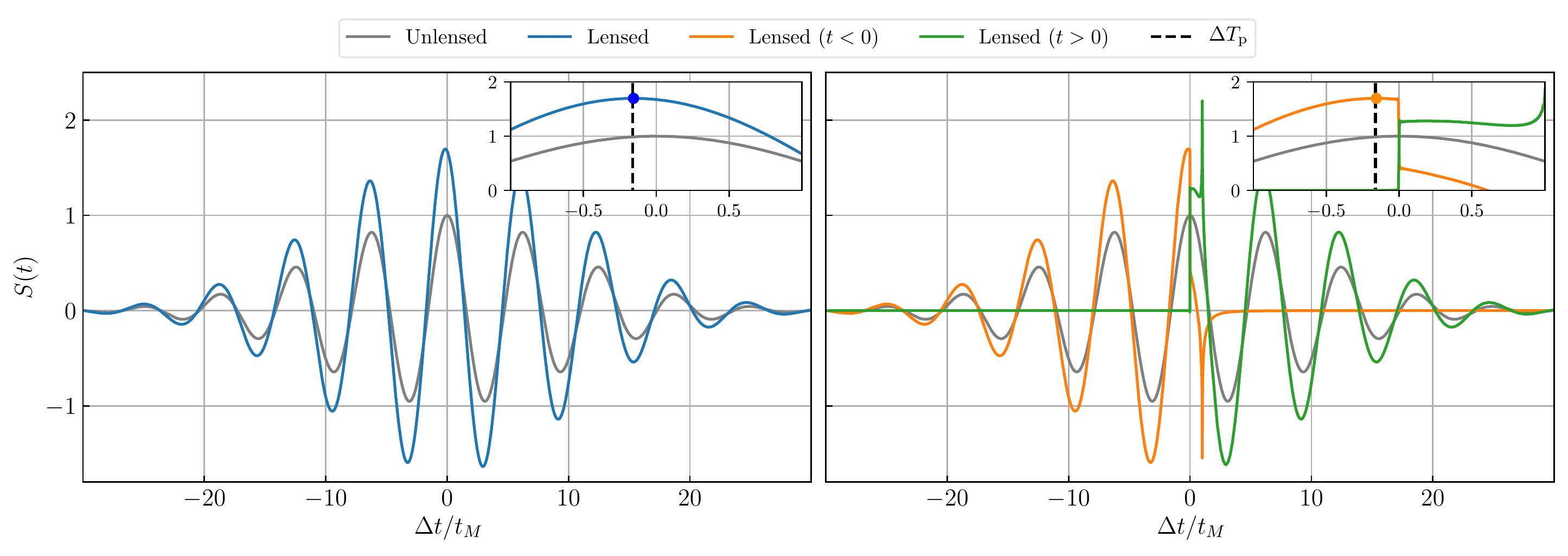}
 \caption{Effect of a point mass lens on a sinusoidal signal modulated by a Gaussian. On the left we plot the lensed signal in arrival time $\Delta t/t_M$, relative
 to the first arrival of light emitted at $t=0$,
 against the unlensed signal (centered so that $\Delta t=0$). Although the peak of the lensed signal arrives before light from $t=0$, corresponding to the advance
 $\Delta T_p$ in the inset (dashed line),
 signals do not propagate superluminally. This can be seen in the right panel where we divide the original signal into two around $t=0$ and show that the division point arrives at the same time as light. 
 The spikes in the half-signals correspond to the secondary images in geometric-optics, arriving at $\Delta T_{-}$.
  In both panels, the carrier frequency $w_0=1$, the dimensionless source position $y=0.5$ and the Gaussian signal width $\sigma=10$.}
 \label{fig:gaussian}
\end{figure*}

For comparison, we plot in Fig.~\ref{fig:phase} the difference between the GW shift $wT_{p,g}$ and EM shift $wT_+$, 
$w \Delta T_{p,g}$ as a function of the frequency, for three different values of the source location.  This highlights the arrival time differences in units of the period of the wave.  Notice that as $w\rightarrow 0$, this quantity always goes to zero from the negative side.

On the other hand, since  both $\Delta T_p$ and $\Delta T_g$ are negative in the low-frequency limit,
the GW phase and group arrival time advances with respect to that of the EM $T_+$ signal, which paradoxically takes the minimum time delay path.  Mathematically, this behaviour can be understood explicitly by considering the limit of small frequencies, $|w|\ll 1$, in which case the GW arrival time becomes independent of the impact parameter \cite{Takahashi:2016jom}:
\begin{align}
    T_\text{p}&\approx \frac{1}{2}\left[\gamma_E+\ln\left(\frac{w}{2}\right)\right]-\ln\theta_E+\mathcal{O}(w\ln w)\,,\\
    T_\text{g}&\approx \frac{1}{2}\left[1+\gamma_E+\ln\left(\frac{w}{2}\right)\right]-\ln \theta_E+\mathcal{O}(w\ln w)\,,
\end{align}
where $\gamma_E$ is the Euler constant. Noticeably, these expressions diverge for $w\rightarrow 0$. The divergence cannot of course represent the actual time delay of a signal since it would indicate that a signal can be received before it is emitted. In fact,
as mentioned above, $\lim_{w\rightarrow 0} F=1$ and the actual waveform is not  distorted by lensing. Instead, it is a mathematical artifact due to the fact that
an infinitesimal phase shift of the GW or
wavepacket center relative to its width can produce
a logarithmically divergent shift in time since
the period or width goes to infinity. 
One can view this as a consequence of the sampling theorem: a wavepacket with only low frequency components can be fully represented by a coarse sampling in time.  In this view, the phase time delay at low frequencies is an uncertainty in the temporal domain rather than a physical delay.

This limiting case already indicates that the phase and group advance should not be viewed as the ability to send GW signals faster than light.  
Indeed, as argued in \cite{Suyama:2020lbf}, this is manifestly apparent in the definition of the amplification $F$: the  geometric optics path  corresponding to $T_+$ is the one that minimizes the travel time and superimposing other paths can only further delay the signal.

From Fig.~\ref{fig:phase} we also see that $T_g>T_p$ and therefore, depending on the lensing parameters, there could be situations where the arrival time of the group  is delayed with respect to light, whereas the arrival time of the phase  is in advance. 
Finally, we notice from Fig.~\ref{fig:phase} that the time delay $\Delta T_i$ is always expected to be a fraction of the period of the wave in the wave optics regime, with a maximum of $w\Delta T_{p,g}\approx 0.6$ (i.e.~a tenth of a period or smaller). 
This limiting case will happen when $y\ll 1$ and $w\sim 1$.
While the phase or group peak can be measured to high precision in principle (see e.g.~\cite{Cremonese:2018cyg}) its interpretation as an advance or delay relative to light requires a matching EM signal as we discuss below.

\section{Apparent Superluminality}\label{sec:gaussian}
In this section, we study an explicit toy model that illustrates the different arrival time behaviors associated with the various meanings of time delay in wave optics and why no information or signal can travel faster than light.  Let us consider a  unlensed signal corresponding to a  single sinusoid modulated by a Gaussian:
\begin{equation}
    S(t)=\cos{(w_0\,t/t_M)}\cdot e^{-\frac{1}{2\sigma^2}\left(\frac{t}{t_M}\right)^2}.
    \label{eq:unlensed}
\end{equation}
Here the  frequency $\omega_0 = w_0/t_M$ carries the phase information, but no temporal localization, while the Gaussian modulation defines the group, with a localized emission centered at $t=0$.

In the left panel of Fig.~\ref{fig:gaussian}, we show the lensed signal compared to the unlensed one as a function of the arrival or emission time, when taking $w_0=1$ (within the wave optics regime), $y=0.5$, and $\sigma=10$. 
In this figure, we have plotted the lensed wave
in dimensionless time $\Delta t/t_M$
relative to the arrival time of light 
emitted at $t=0$ in the $T_+$ geometric optic limit.
For the 
unlensed wave, we correspondingly plot Eq.~(\ref{eq:unlensed}) with $t \rightarrow \Delta t$ so that the central peak of the unlensed signal corresponds to the lensed wave with a time delay relative to light of $\Delta T=0$.
 Here we explicitly see that the central peak  of the signal arrives before light ($\Delta t<0$), leading to the apparent interpretation that GWs in the wave optics regime propagate faster than light. The inset of the 
left panel zooms into the region  around the maximum peak (indicated with a blue dot). The time delay of this maximum point is well approximated by the phase time delay $\Delta T_p$ (indicated in the inset with a vertical black dashed line).

However, this maximum point in the lensed waveform 
does not correspond to emission from the maximum of the
unlensed waveform as one can confirm by dividing the signal into two parts:
\begin{align}
  S(t)&= S_1(t) + S_2(t);  \\   S_1(t)&=[1-\Theta(t)]S(t), \, S_2(t)=\Theta(t)S(t),
  \nonumber
\end{align}
where $\Theta(t)$ is the Heaviside step function.
Linearity of the wave equation guarantees that the superposition of the two lensed waveforms gives the
total lensed waveform as well.

In this case, both signals around the step are dominated by high frequency modes and hence we expect their discontinuous edge to travel at the front velocity and arrive at the same time as light. This is indeed what we see on the right panel of Fig.~\ref{fig:gaussian}, where both lensed signals (orange and green curves) exhibit a step at $\Delta t=0$. Notice that the trailing half is strictly zero for $\Delta t<0$ whereas the leading half produces contributions for $\Delta t>0$.  This asymmetry is in accordance with causality and the additional delay due to wave optics paths away from the minimum of the time delay $T_+$.
The sum produces the illusion of a signal propagating superluminally since its maximum occurs at $\Delta t<0$.  The same logic applies to nodes of the unlensed waveform.  The finite amplitude of the lensed waveform
at the corresponding light arrival time and the apparent superluminality that it implies 
actually arises from the enhanced time delay of emission from an earlier, not later, epoch.

Also note that in the right panel Fig.~\ref{fig:gaussian}, the sharp spike  of the lensed $S_1$ and $S_2$ signals for $\Delta T>0$ correspond to the second image of the geometric optics regime and hence they arrive at $\Delta t/t_M= \Delta T_{-}$.  These of course cancel when the two pieces are superimposed. 
The fact that second image is fainter can be seen by comparing the relative amplitude of the step in $S_2$ at $\Delta T_+\equiv 0$ to the spike at $\Delta T_-$ above the rest of the wave.

\begin{figure}[t!]
\centering
\includegraphics[width = 0.95\columnwidth]{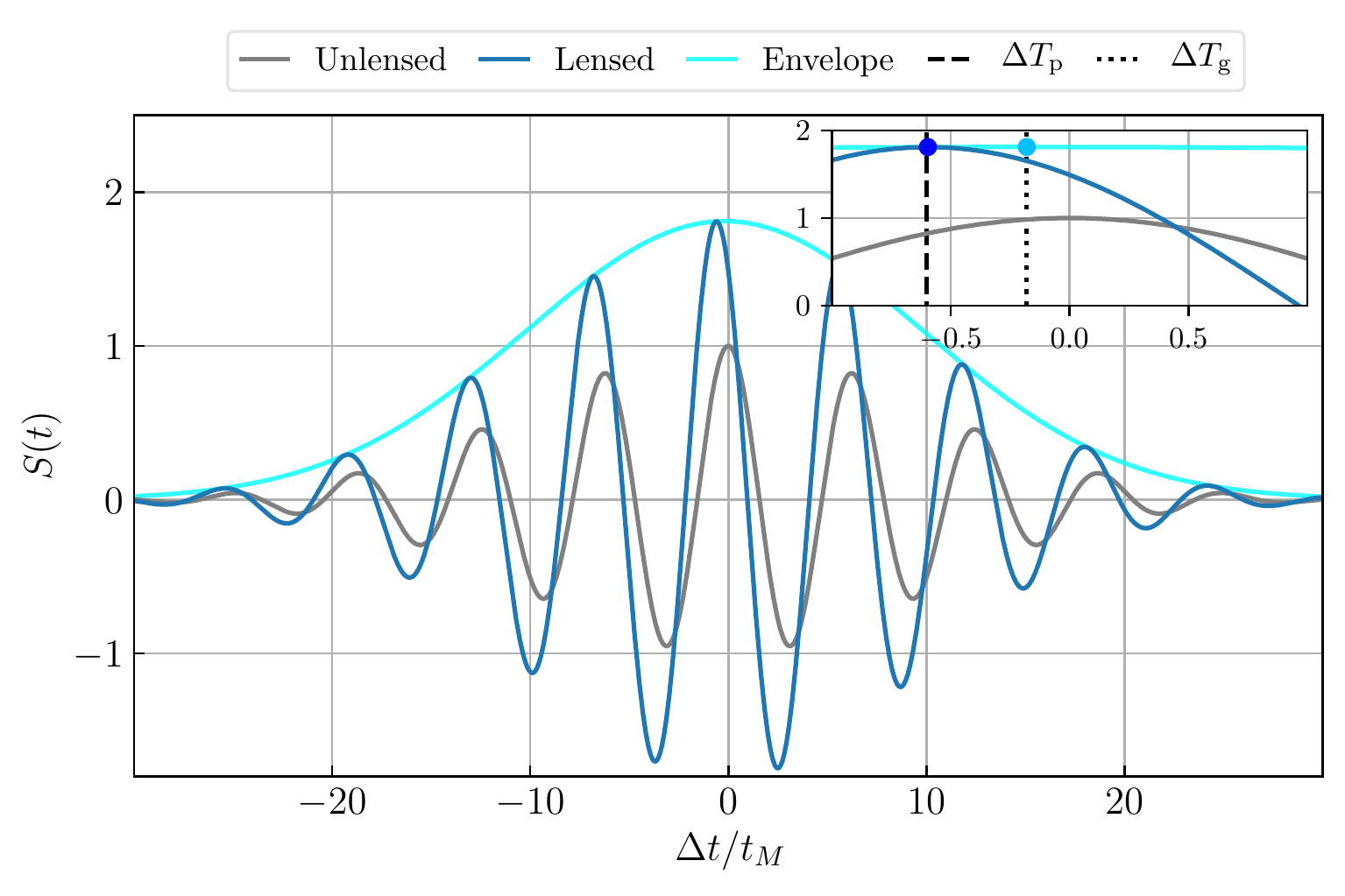}
 \caption{Envelope (cyan) of lensed modulated sinusoid signal (blue). 
  The maximum of envelope (cyan dot in inset) arrives at $\Delta T_g$, the group delay with respect to light (black dotted line), whereas the maximum of the central peak (blue dot in inset) arrives at $\Delta T_p$, the phase delay  (black dashed line).
 Here $w_0=1$, $y=0$, $\sigma=10$.}
 \label{fig:group_velocity}
\end{figure}

That the phase delay carries no temporally localized signal and does not indicate superluminality is a familiar concept from EM waves in a plasma.   However in the GW case, the group delay can also show an apparent superluminality.
The temporal localization of the unlensed waveform is
characterized by the Gaussian envelope and causes 
$S_\omega$ to have a Gaussian spread of frequencies
around $w_0$.
In Fig.~\ref{fig:group_velocity} we show 
this envelope in the lensed waveform.  

The maximum of this envelope can be delayed or advanced with respect to light, depending on the lensing parameters. For the values in Fig.~\ref{fig:group_velocity}, the maximum of the envelope also has an apparent arrival before light. This is explicitly seen in the inset of Fig.~\ref{fig:group_velocity}, where the maximum of the envelope is indicated by a cyan dot, and is seen to arrive at $\Delta T_g<0$ (indicated by the vertical black dotted line).  
However, as demonstrated in Fig.~\ref{fig:gaussian}, this advance in the arrival of the group peak again 
reflects a distortion of the waveform due to the net lensing amplification of multiple paths arriving to the observer as well as enhanced delay with respect to light rather than information that propagates superluminally. 

\begin{figure*}[t!]
\centering
\includegraphics[width = 0.98\textwidth]{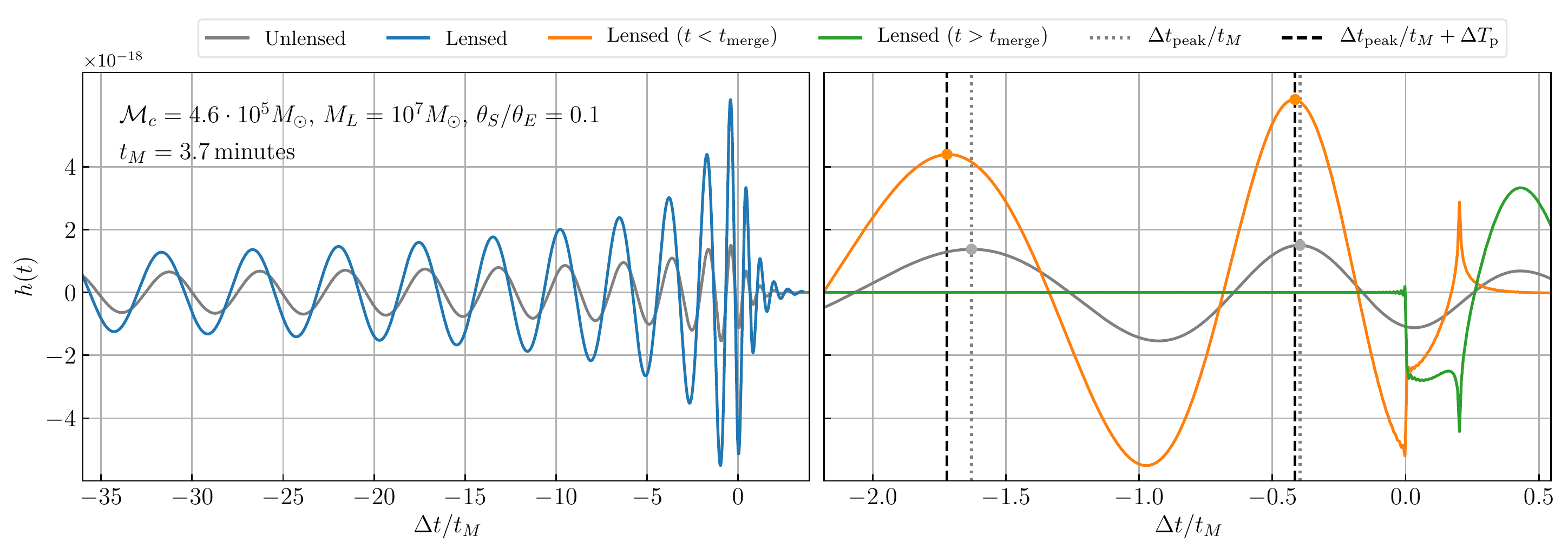}
 \caption{Effect of a point mass lens on a GW from a compact binary coalescence. We fix $\Delta t=0$ with the merger time $t_\mathrm{merge}$ and the first arrival of light from this time in geometric optics $t_+$, for the unlensed and lensed signals, respectively. In the left panel we show the total lensed (blue) and unlensed (black) wave, whereas in the right panel we also show the lensed signal divided into two parts: $t<t_\mathrm{merge}$ and $t>t_\mathrm{merge}$. The peaks of the waveform are delayed by $\Delta T_p$, the phase delay, whereas the merger front arrives with light.
 The GW is always in the wave-optics regime given that we have chosen $m_1=10^6 M_\odot$, $m_2=3\cdot10^5 M_\odot$, $z_s=1$, $M_L=10^7 M_\odot$ and $z_L=0.1$.  Since $t_M=3.7\,$minutes,  on the left we are observing the last 2 hours before merger and on the right we zoom-in on the last 6 minutes. Note that the small oscillations and deviations from a discontinuous step around $t_\mathrm{merge}$ on the right panel are just a numerical artifact of the finite maximum frequency of the Fourier transform employed.
 }
 \label{fig:gw}
\end{figure*}

\section{Lensed Binary Coalescence Waveform}\label{sec:waveform}

With an understanding of the meaning of various time delays from the simple wave-packet example of the previous section, we now turn to a realistic waveform from a binary black hole and show its lensed form from the inspiral to the ringdown. 

Let us first consider when this waveform is in the wave optics limit.
The typical frequency of a compact binary merger is given by the inner most stable circular orbit ($R\simeq3R_S$) implying a redshifted observed frequency $f_\mathrm{gw}\simeq220(1+z_S)^{-1}(20M_\odot/M_S)$Hz, for a source of mass $M_S$ at redshift $z_S$. Accordingly, wave-optics effects will be relevant whenever
\begin{equation}
    (1+z_L) M_L \lesssim 10\,(1+z_S) M_S\,,
\end{equation}
where $z_L$ and $M_L$ correspond to the redshift and mass of the lens, respectively.
Alternatively, directly in terms of the observed frequency, the condition reads
\begin{equation}\label{fgwWaveOptics}
    f_\text{gw}\lesssim 5\cdot{10^{4}}(1+z_L)^{-1} \left(\frac{M_\odot}{M_L}\right)\text{Hz}\,.
\end{equation}
This means that in the ground-based detector band, $5-5000$Hz, wave-optics is relevant for lenses between $10-10^3M_\odot$ or smaller. Similarly, in the band of space-based detectors such as LISA, $10^{-4}-10^{-1}$Hz, the lenses would be $10^{5}-10^{8}M_\odot$ or smaller. For PTA, the observed frequencies range between $10^{-9}-10^{-6}$Hz, and thus wave optics would be relevant for lens masses $10^{10}-10^{13}M_\odot$ or smaller.

As an example where both GW and EM emissions might be expected during the inspiral phase (e.g.~\cite{2012MNRAS.420..860S, Tanaka:2013oju, Haiman:2017szj, DalCanton:2019wsr}),
we consider as a concrete case of a supermassive binary black hole, with masses $m_1=10^6\,M_\odot$ and $m_2=3\cdot10^5\,M_\odot$, hence a mass ratio $q=0.3$. However, our results can be directly extrapolated to other ranges of masses by rescaling appropriately the amplitudes and times. Fig.~\ref{fig:gw} shows the waveform from inspiral to ringdown.
We further assume a source at redshift $z_S=1$ with a lens of mass $M_L=10^7M_\odot$ at $z_L=0.1$ and angular position $y=0.1$.
We transform redshift to distances assuming a $\Lambda$CDM cosmology with Planck's 2018 parameters \cite{Collaboration:2018aa}.
We ignore spin effects and use IMRPhenomD waveform model \cite{PhysRevD.93.044006} implemented in pyCBC \cite{alex_nitz_2019_3546372}.

We make an analogous analysis to the previous section, and show in the left panel of Fig.~\ref{fig:gw} the total lensed vs unlensed signal, and in the right panel the signal divided in two parts at the moment of the merger $t=t_\mathrm{merge}$. We use the same normalization for time delays as the previous section. 
We can see that the peaks of the lensed GW (blue line) during the inspiral all indeed arrive before the corresponding peaks of the unlensed signal (black line). 

In this example, wave optics is valid throughout the entire GW emission, with the dimensionless frequency $w$ varying from $\sim 1$ to $2\pi$ in the time span of Fig.~\ref{fig:gw}.
Since at any given time, the GW emitted is monochromatic, we expect the peaks of the GW signal to have time delays well approximated by the phase time delay $T_p$. This is indeed what can be seen in the right panel. The maximum of the two last peaks before the merger are indicated with orange dots for the lensed wave and grey dots for the unlensed one. The time difference is given by $T_p$ (evaluated at the observed frequencies of the GW signal at the time of arrival of the last two unlensed peaks before the merger), as shown with black vertical dashed lines. In this case, while the chirp does carry a signal of the merger, the phase is arbitrary and cannot temporally localize the merger event exactly at $t=t_\mathrm{merge}$. 

The envelope of the chirp profile does have this merger time information in principle, and the delay of its peak is characterized by the
group delay $T_g$ at some effective frequency, as long as all frequencies of the signal are in the wave optics regime. However, as shown in the previous section, if the
envelope is wide compared to $t_M$,
the peak of the lensed envelope cannot be used to determine a precise merger time. In these cases, an advance of this envelope peak relative to light does not reflect superluminal propagation of the merger event.

We can again explicitly verify this by dividing the gravitational wave signal into two at the merger event $t=t_\mathrm{merge}$. As in our example of a modulated sinusoid, the signal coming from $t=t_\mathrm{merge}$ is represented by the steps in the
two half signals and arrives simultaneously with light.

\section{multi-messenger Discussion}\label{sec:discussion}

In this paper, we analyze the time delay between simultaneously emitted GW and EM signals, in the presence of lensing. For astrophysical lenses, the lensing of EM signals is well approximated by the geometric optics regime. However, GW observations
can have wavelengths that are greater than the Schwarzschild radius of the lens,
in which case lensing must be treated in the wave optics regime, where interference effects will become relevant. For these scenarios, we clarify the meaning of different definitions of time delay based on the phase, group, and front velocity of GW. We find that the GW features determined by the phase or group evolution of the signal may appear to arrive in advance to its corresponding simultaneously emitted EM signal, in an apparent contradiction with causality as light in the geometric optics regime takes the shortest path from the source to the observer. We argue that this paradox is resolved by noticing that interference distorts the GW signal and effectively leads to peaks of the signal to be in advance. However, we confirm with concrete examples that the front velocity of GW does coincide with that of light, and therefore there is no superluminal propagation of GW. 

Even though the apparent superluminality of the phase delay does not represent a signal that actually propagates superluminally, there can still be observational consequences if there is an EM counterpart that also carries this phase information. 
For supermassive black hole (SMBH) binaries, there
can be EM counterparts with features that evolve with the binary's orbit and so  are synchronized with the GW signal in phase. Numerical simulations show that SMBH binaries in a gaseous environment develop a central circumbinary cavity surrounded by an accretion disk \cite{1994ApJ...421..651A, 2005ApJ...622L..93M}. This circumbinary gas will periodically leak into the cavity and create disks bound to the individual SMBHs, called minidisks \cite{2008ApJ...682.1134H}. It is still unclear how long these inner disks live, but if they live at least for many orbital cycles, then the EM emission from these minidisks will inevitably have features due to relativistic Doppler effects which will track the binary's motion. This Doppler effect is expected to be more dominant for unequal binary mass ratios \cite{DOrazio:2015shf}, like the one we present in Fig.~\ref{fig:gw}. 
The EM periodic features could be associated to periodic changes both in the spectra \cite{2012MNRAS.420..860S, McKernan:2013cha} (e.g.~shifts in a given emission/absorption line) or brightness \cite{Haiman:2017szj, DalCanton:2019wsr}. Another potential scenario has been proposed in \cite{Zalamea:2010mv} for binaries with a supermassive BH and a white dwarf (WD). In that case, the WD may be tidally disrupted and exhibit periodic gentle mass losses for thousands of orbital periods, potentially creating a relatively long-lived source of gas that accretes into the black hole and produces a periodically varying EM radiation tracing the binary's motion for eccentric orbits. Another proposal for WD binaries has been discussed in \cite{Cooray:2003qm, Lamberts:2019nyk} where the prolonged timing of eclipses could also be used to track the binary's orbital motion and period evolution.

Having an EM signal that is synchronized with the GW inspiral phase has several important implications. First of all, a multi-messenger \emph{pre-merger} observation can be used to localize the event \cite{DalCanton:2019wsr}, dramatically improving the chances of detecting any \emph{post-merger} EM transient. Moreover, when the signals are lensed, the phase delay before the merger could inform on the time delays between the possible post-merger multiple images. 
Measuring these phase delays could also provide an independent constraint on the lens mass model
by observing its frequency evolution over many cycles, as well as on cosmological parameters such as $H_0$ \cite{Cremonese:2019tgb}.

In addition, the phase delay can constrain the speed of GWs, which is a fundamental property of any gravity theory. 
Forecasts for constraints on the GW propagation speed and graviton mass from  LISA can be found in \cite{Haiman:2017szj} for SMBHs and in \cite{PhysRevD.88.022001,Bettoni:2016mij} for LISA verification binaries. For instance, constraints on the graviton mass are expected to improve by 4 orders of magnitude the current LIGO constraints from waveforms. Therefore, knowing if the GW is lensed is essential in order to correctly interpret if the delay has an astrophysical or fundamental origin.

The probability of a given source to be lensed by a galaxy depends on its distance, lens distribution, and lensing cross section (determined by the type of galaxy) \cite{1984ApJ...284....1T}. For instance, for a source at redshift $z\sim 1$ the probability is estimated to be of order $10^{-1}-10^{-3}$. Realistically, whether GW detectors will see lensing in the wave optics regime will also depend on the population of compact binaries and experimental sensitivities. Initial studies for PTA have been made in \cite{Takahashi:2016jom, Cremonese:2018cyg}, concluding that if a  hundred of events can be resolved by PTA in the SKA era \cite{2012MNRAS.420..860S} then some lensed events will be detected. Even if only one of these events is detected, it may be used to set competitive constraints on cosmological parameters such as $H_0$ \cite{Cremonese:2019tgb}. 
Regarding LISA, different formation channels for SMBH binaries were studied in \cite{PhysRevLett.105.251101}, concluding that LISA might detect up to several multiple-image lensed events in the geometric optics regime. Moving forward, diffraction effects could be identified searching for modulations in the amplitude and phase of the wave-forms \cite{PhysRevD.98.104029}. 
Alternatively, as we have seen here, multi-messenger measurements of the phase lag between GWs and EM radiation can be used to identify lensed events in wave-optics limit.

Looking to the future, more detailed analyses on how well the EM counterparts can be uniquely identified to a host at high redshifts, and how well the EM phase due to Doppler motion can be measured will be needed to fully assess the viability of using time delays during the inspiral of binaries to test gravity and cosmology. 
Moreover, realistic predictions also require the modelling of the effect of surrounding matter to the lens, specially for microlenses embedded in macrolenses \cite{refId0}.
Nevertheless, the multi-messenger time delays described in this work show potential for extracting independent observational constraints, and our results help interpret appropriately the physics of the signals received.

\acknowledgements
We are grateful to Zoltan Haiman, Daniel Holz,  Meng-Xiang Lin, and Sam Passaglia for useful conversations. 
JME is supported by NASA through the NASA Hubble Fellowship grant HST-HF2-51435.001-A awarded by the Space Telescope Science Institute, which is operated by the Association of Universities for Research in Astronomy, Inc., for NASA, under contract NAS5-26555. 
WH was supported by the U.S.~Dept.~of  Energy contract DE-FG02-13ER41958 and the Simons Foundation. 
ML was supported by the Kavli Institute for Cosmological Physics at the University of Chicago through an endowment from the Kavli Foundation and its founder Fred Kavli. 

\vfill
\bibliography{RefModifiedGravity}

 \end{document}